\documentclass{kluwer}    
\usepackage{epsfig}
\def\ts{\thinspace}
\def\um{$\mu$m}
\def\etal{{\it et al.}}

\newdisplay{guess}{Conjecture}
\begin{document}                                                                
                  
\begin{article}
\begin{opening}         
\title{A New View of Galaxy Evolution from Submillimeter Surveys with SCUBA}
\author{D. B. \surname{Sanders}}  
\runningauthor{D. B. Sanders}
\runningtitle{Submm Surveys with SCUBA}
\institute{Institute for Astronomy, University of Hawaii, 2680 Woodlawn Drive,
Honolulu, HI\ 96822, \email{sanders@ifa.hawaii.edu}}

\begin{abstract}
Our view of galaxy evolution has been dramatically enhanced by the recent deep
field submm surveys carried out with the SCUBA camera on the JCMT.
SCUBA has discovered a population of luminous infrared galaxies at redshifts
$\sim${\ts}1--4 that emit most of their energy at far-IR/submm 
wavelengths.  The cumulative surface density of submm 
sources ($\sim${\ts}10$^4${\ts}deg$^{-2}$ with 
$S_{850}${\ts}$>${\ts}1{\ts}mJy) appears
to be sufficient to account for nearly {\it all} of the 850{\ts}$\mu$m
extragalactic background.  The SCUBA sources are plausibly the
high-$z$ counterparts of more local ($z${\ts}$\lsim${\ts}1) 
luminous infrared galaxies 
that have been identified in {\it IRAS} and {\it ISO}
deep field surveys, the majority of which appear to be major mergers of gas-rich
disks accompanied by dust-enshrouded nuclear starbursts and powerful AGN. 
The SCUBA sources
are plausibly the progenitors of the present-day spheroidal population.  This
major event in galaxy evolution, equal in bolometric luminosity to that
observed at optical wavelengths, is largely missed by current UV/optical
surveys.  
\end{abstract} 


\end{opening}           

\vspace{-0.25cm}
\section{Introduction}  
The Submillimeter Common User Bolometer Array (SCUBA) camera
on the James Clerk Maxwell
Telescope (JCMT) (Holland \etal\  1999) has provided a new window for
ground-based
studies of the high-$z$ Universe.  This brief review summarizes 
results from a large campaign of deep 
surveys carried out during SCUBA's first two years of operation on
Mauna Kea.  Evidence is presented that the submm sources detected in the SCUBA 
deep fields must be predominantly luminous infrared galaxies 
(LIGs: $L_{\rm ir}${\ts}$>${\ts}10$^{11}${\ts}$L_\odot$)\footnote{$L_{\rm
ir}${\ts}$\equiv${\ts}$L$(8--1000$\mu$m).  
Unless otherwise stated, 
$H_{\rm o}${\ts}$=${\ts}50{\ts}km{\ts}s$^{-1}$Mpc$^{-1}$,
$q_{\rm o}${\ts}$=${\ts}0.} at high redshift 
($z{\ts}\sim${\ts}1--5), and by analogy with local LIGs, 
that they plausibly represent the building of 
spheroids through major mergers of gas-rich disks. 

\vspace{-0.25cm}
\section{SCUBA Deep Surveys}
 
Smail \etal\  (1997)
were the first to infer a substantial population of luminous
submm galaxies from their SCUBA detections at
850{\ts}$\mu$m/450{\ts}$\mu$m of background sources amplified
by weak lensing from foreground clusters.  
Subsequent blank-field 
surveys at 850\um/450\um\ (Hughes
\etal\  1998; Barger \etal\  1998; 
Eales \etal\  1999), confirmed the surprisingly large space density of faint
submm
sources, and in addition, showed that their optical 
{\it and near-infrared} counterparts were often quite faint, as 
illustrated below in our own data for the Lockman Hole. 

\subsection{The Lockman Hole}
 
The 850{\ts}$\mu$m data for the Lockman Hole deep field 
are shown in Figure 1.  The two SCUBA
sources, LH\_NW1 and LH\_NW2, detected at 850{\ts}$\mu$m
($>${\ts}3{\ts}$\sigma$), have 850{\ts}$\mu$m fluxes of 5.1{\ts}mJy,
and 2.7{\ts}mJy, respectively, with upper limits at 450{\ts}$\mu$m
of $\lsim${\ts}50{\ts}mJy (5{\ts}$\sigma$).
Neither SCUBA source has an ISOCAM 7{\ts}$\mu$m counterpart
($< 35${\ts}$\mu$Jy; 5{\ts}$\sigma$).  LH\_NW1 appears to be
centered on a faint K$^\prime$ source ($K_{\rm AB}^\prime = 21.8$)
with disturbed morphology, which is barely detected in the current
B-band image ($B_{\rm AB} = 23.5$).  LH\_NW2 is ``blank" 
implying that any counterpart has $K_{\rm AB}^\prime > 22.5$
and $B_{\rm AB} > 24.5$.  

\begin{figure}[hb]
\vspace{-0.5cm}
\centerline{\epsfig{file=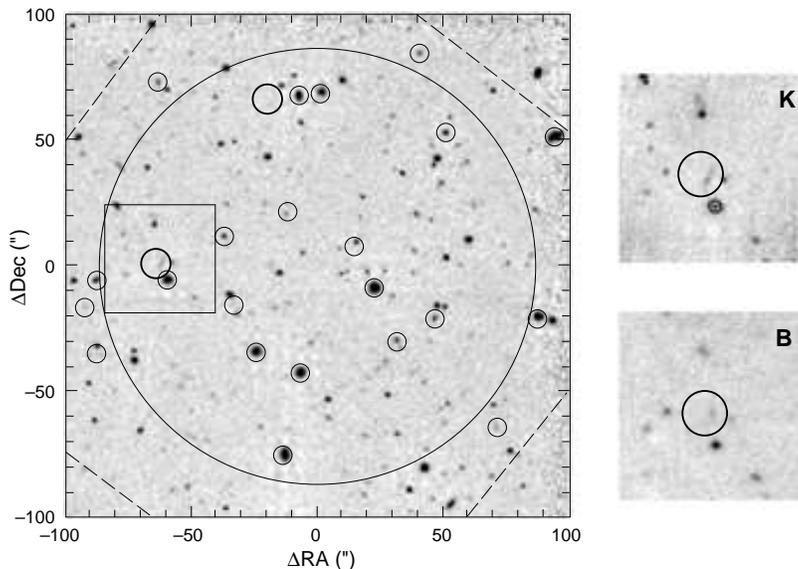, width=12cm}}
\caption{SCUBA 850{\ts}$\mu$m detections (2 small thick circles: Barger \etal
\
1998; Barger, Cowie \& Sanders 1999), and ISOCAM 7$\mu$m detections (22 small
thin circles: Taniguchi \etal\  1997) in the Lockman Hole northwest
(LH\_NW) Deep Field
(J2000: $RA = 10^h33^m55.5^s, Dec = +57^\circ46^\prime18^{\prime\prime}$)
superimposed on a K$^\prime$ image obtained with the QUick InfraRed
Camera (QUIRC) on the University of Hawaii 2.2-m telescope.  The
field-of-view of the ISOCAM detector array and the SCUBA array are
indicated by a long dashed line and large solid circle respectively.
On the right are two zoomed images of the region outlined by the
$45^{\prime\prime} \times 45^{\prime\prime}$ box which is centered
on the strongest SCUBA source.  The zoomed K$^\prime$ image was
obtained with the Near InfraRed Camera (NIRC) on the Keck 10-m
telescope, and the zoomed B-band image was obtained with the University
of Hawaii 2.2-m telescope.}
\vspace{-0.5cm}
\end{figure}

\section{ULIGs at High Redshift}

\begin{figure}[hb]
\vspace{-0.25cm}
\centerline{\epsfig{file=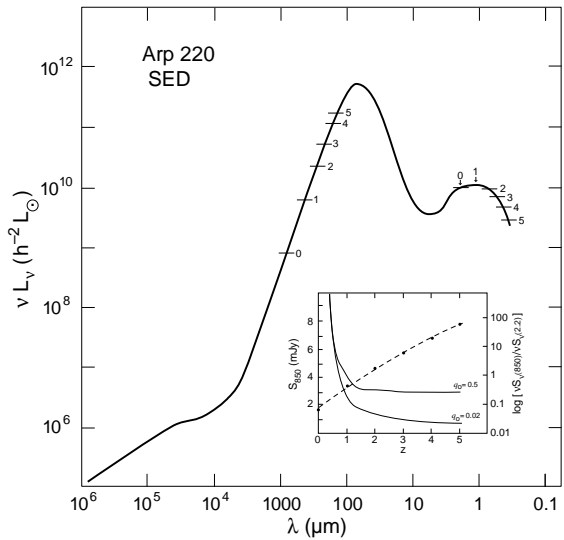, width=7.5cm}}
\caption{Observed radio-to-UV spectral energy distribution of the nearest 
ULIG,
Arp{\ts}220 ($z = 0.018$).  Labeled tickmarks represent object rest-frame 
emission
that will be shifted into the 850{\ts}$\mu$m and 2.2{\ts}$\mu$m
observed frame for redshifts, $z = 0 - 5$.  The insert shows the
corresponding observed-frame 850{\ts}$\mu$m flux and $\nu S_\nu(850)/\nu S_
\nu(2.2)$
ratio for Arp{\ts}220 at redshifts, $z = 0 - 5$.
}
\end{figure}

From the strength of the 850{\ts}$\mu$m detections and the faintness
of the K$^\prime$ counterparts alone, it is relatively straightforward
to show that the SCUBA sources detected in the LH\_NW deep field are most 
likely to be ultraluminous infrared galaxies 
(ULIGs: $L_{\rm ir}${\ts}$>${\ts}$10^{12}${\ts}$L_\odot$) at high redshift
(i.e. $z >${\ts}1).  The ``submm excess",
($\equiv \nu S_\nu (850\mu m) / \nu S_\nu (2.2\mu m)$), for both LH\_NW1 and
LH\_NW2
is larger than 1 (2.4 and $>${\ts}3 respectively), which is impossible to
produce 
with normal optically selected galaxies at any redshift, or even by the
most extreme infrared selected galaxies at low redshift, but is almost
exactly what would be expected for an ULIG at high redshift.  Figure 2
shows that the expected flux for the nearest ULIG Arp{\ts}220 when placed at
$z >${\ts}1 is on the order of a few mJy at 850{\ts}$\mu$m. Also, the 
combination of a
large negative K-correction in the submm plus a
relatively flat or positive K-correction in the near-IR
naturally leads to values $\nu S_\nu (850\mu m) / \nu S_\nu (2.2\mu m) > 1$ for
all ULIGs at $z \gsim${\ts}1.5{\ts}.  The observed faintness of the high-$z$
submm sources
in current B-band images and the non-detections at 7{\ts}$\mu$m in the deep
ISOCAM images are consistent with the large U--B colors and the
pronounced minimum at $\sim${\ts}3-6{\ts}$\mu$m respectively, in the 
rest-frame SEDs of ULIGs
like Arp{\ts}220.  

\section{Source Counts, the Extragalactic Background, and Luminosity Function
Evolution}

\begin{figure}[hb]
\centerline{\epsfig{file=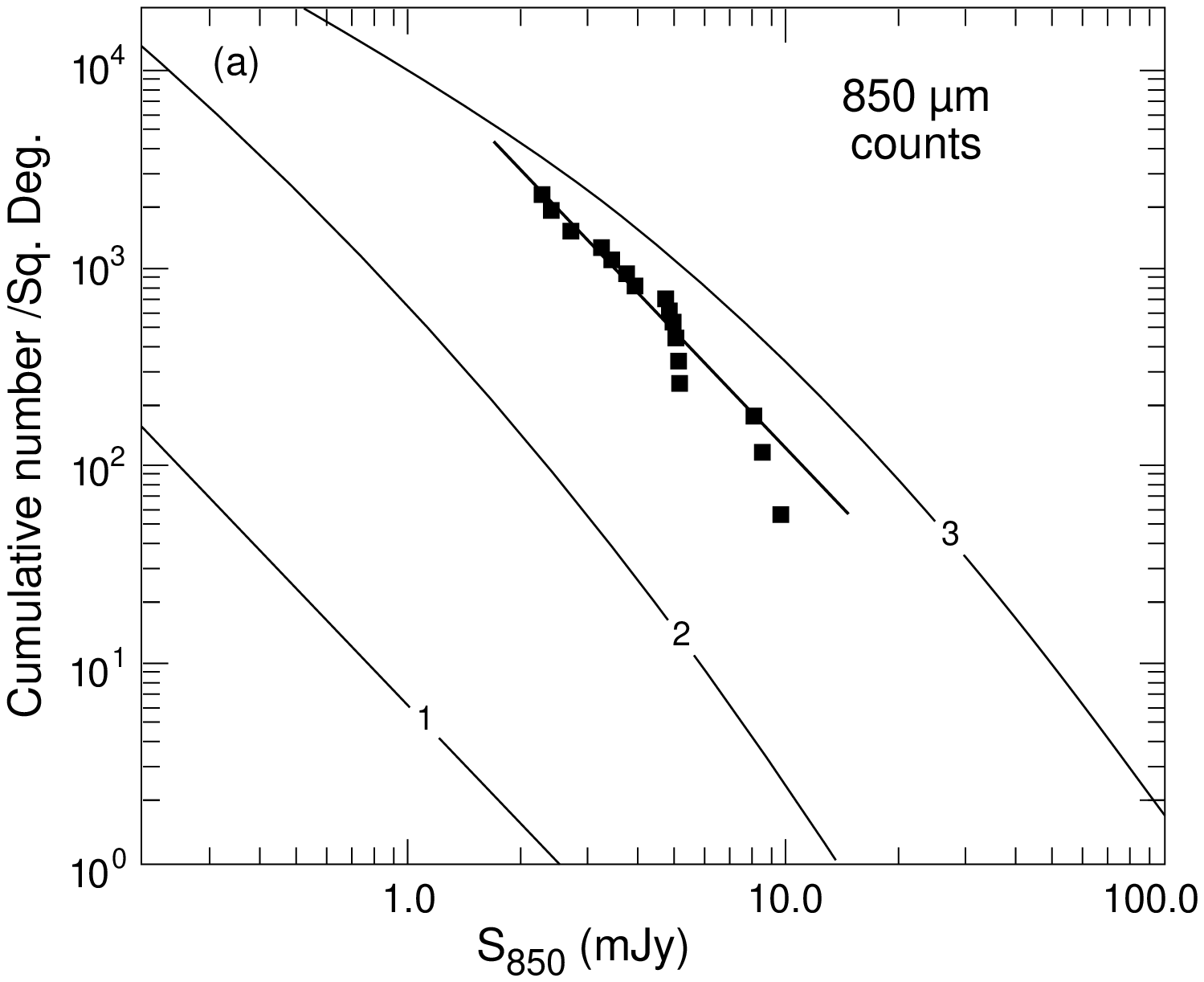, width=6cm}\ 
\epsfig{file=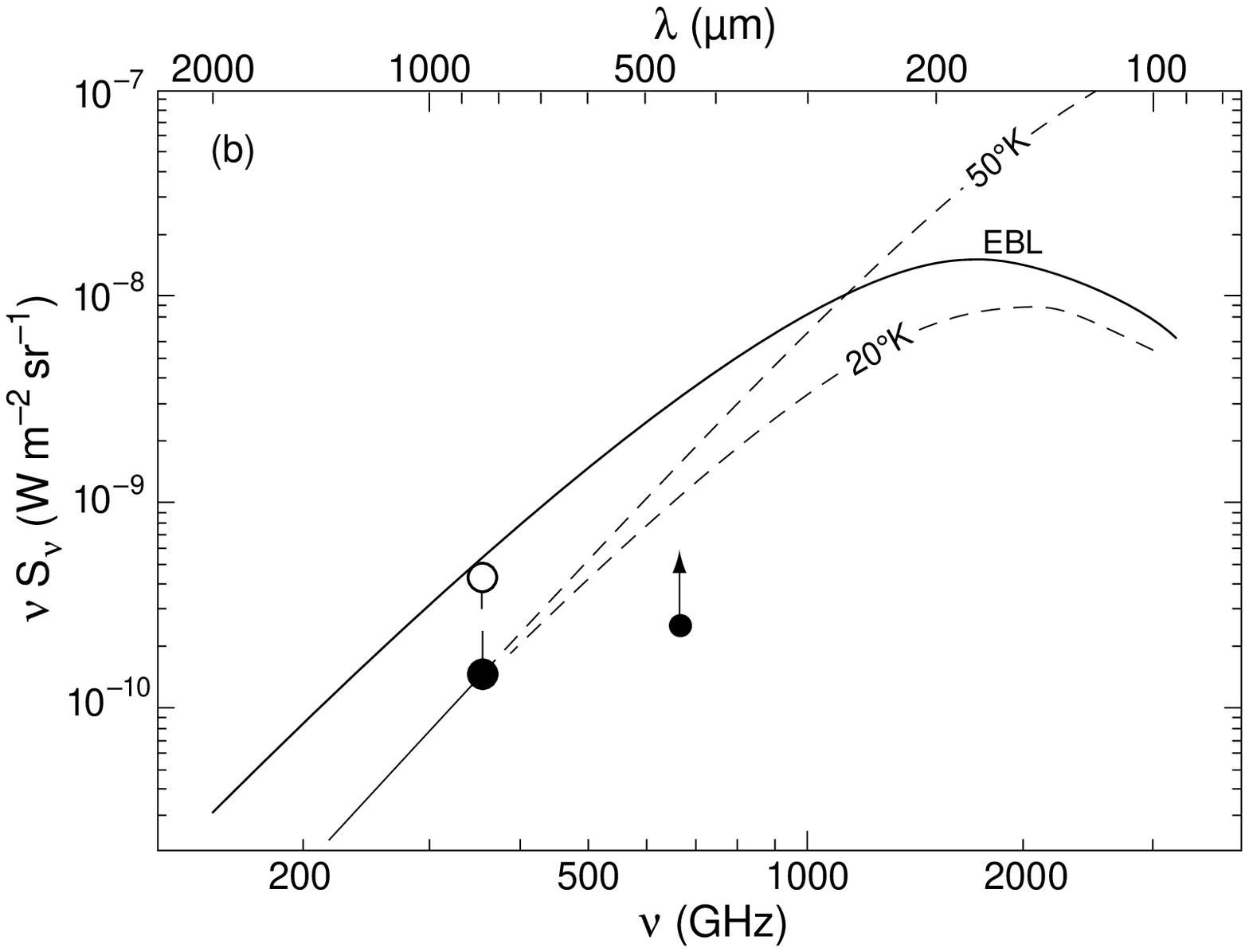, width=6cm}}
\caption{{\it (a)}\ Comparison of the 850{\ts}$\mu$m source counts
(solid squares: from Barger, Cowie \& Sanders  1999) with semi-analytic model
counts (see text).  
 \ {\it (b)}\ 
Comparison of the contribution of the 850{\ts}$\mu$m sources
brighter than 3{\ts}mJy (solid circle) and extrapolated contribution
of sources brighter than 1{\ts}mJy (open circle) to the EBL compared
with the Fixsen \etal\  (1998) analytic approximation (solid curve) to the
EBL.
The two dashed curves are for {\it observed} source temperatures
of 50{\ts}K and 25{\ts}K where each is based on a $\lambda$-weighted
Planck function.}
\end{figure}
 
Figure 3a shows that the cumulative 850{\ts}$\mu$m
counts in the range 2--10{\ts}mJy can be approximated by a single
power law of the form $N(>S) = 1 \times 10^4\ S^{-2}$ deg$^{-2}$.
Figure 3b compares the contribution of these 850{\ts}$\mu$m sources with
the recent model of the EBL determined from COBE data 
(Fixsen \etal\  1998; Puget \etal\ 1996, 1999; 
Hauser \etal\  1998).  Approximately 25{\ts}\% of the 
850{\ts}$\mu$m EBL resides in sources brighter than 2{\ts}mJy,
and {\it nearly all of the EBL at 850{\ts}$\mu$m can be accounted
for by sources brighter than 1{\ts}mJy}, assuming the extrapolation down to
1{\ts}mJy given by the fit to the SCUBA data in Figure 3a.

The observed cumulative SCUBA counts imply {\it strong evolution} in 
the co-moving space density of LIGs and ULIGs. 
Figure 3a compares the observed SCUBA counts with
predictions from semi-analytic models using three rather extreme 
distributions of ULIGs.
Model 1 is based on the local IRAS 60{\ts}$\mu$m luminosity
function of galaxies (i.e. $\sim${\ts}0.001 ULIGs deg$^{-2}$ at
$z${\ts}$<${\ts}0.08: 
Kim \& Sanders 1998; see also Soifer \etal\  1987; Saunders \etal\  1990)
{\it assuming no evolution, which underestimates the observed space 
density SCUBA sources by
nearly 3 orders of magnitude}.   Model 2 includes no ULIGs, instead
attempting to account for the fraction of the optical/UV emission absorbed
and reradiated by dust in sources observed in optical/UV deep fields.
Model 2 still underpredicts the 850{\ts}$\mu$m source
counts by a factor of $\sim${\ts}30.  A better fit to the data is provided 
by Model 3 (similar to Model E of Guiderdoni \etal\ 1998), which 
includes a strongly evolving
population of ULIGs, constrained only by recent measurements
of the submm extragalactic background light (EBL).
and 

Figure 4 graphically illustrates how the high luminosity tail of the 
LF for infrared galaxies 
must change to match the observed SCUBA counts and inferred redshift 
distribution (Barger et al. 1999).
It is interesting to note that 
the strong evolution already detected in the 1-Jy sample of ULIGs
over the relative small range 
$z${\ts}$\lsim${\ts}0.3 [i.e. $\propto (1+z)^{6-7}$: Kim \& Sanders 1998], 
if continued out 
to $z${\ts}$\sim${\ts}2, would also provide a good 
match to the observed cumulative surface density of SCUBA sources. 

\begin{figure}[hb]
\centerline{\epsfig{file=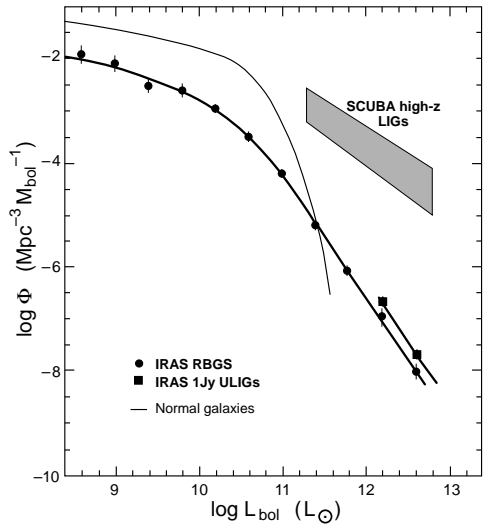, width=6.5cm}}
\caption{The local LFs for infrared selected galaxies 
from the {\it IRAS\/} Bright Galaxy Sample 
(Soifer \& Neugebauer 1991; Sanders \& Mirabel 1996)
and for optically selected ``normal'' galaxies (Schechter 1976), 
compared to the LFs for slightly more distant ULIGs
from the {\it IRAS\/} 1-Jy sample (Kim \& Sanders 1998) 
and for the high-$z$ submm sources detected in the SCUBA 850{\ts}$\mu$m 
deep fields.}
\end{figure}

\vspace{-0.25cm}
\section{Identification of IR/Submm Sources}
\vspace{-0.25cm}

\subsection{Low-z LIGs ($z${\ts}$\lsim${\ts}0.3)}
Substantial progress has been made in understanding the nature of infrared
selected 
galaxies in the local Universe.  Ground-based follow-up studies of complete
samples 
of LIGs discovered by the {\it IRAS} satellite show that 
nearly all objects with $L_{\rm ir} > 10^{11.5}${\ts}$L_\odot$ appear to be 
strongly interacting/merging, gas-rich,  
$\sim${\ts}$L^*$ spirals (Figure 5). 
At the highest luminosities most objects 
appear to be advanced mergers powered by a mixture of starburst and AGN both of
which are 
fueled by an enormous concentration of gas that has been funneled 
into the merger nucleus.  These LIGs appear to represent a primary stage in the
formation 
of elliptical galaxy cores, and  the ULIG phase also appears to represent an 
important phase in the 
formation of quasars and powerful radio galaxies (see SM96 for a 
complete review).  

\begin{figure}[hb]
\centerline{\epsfig{file=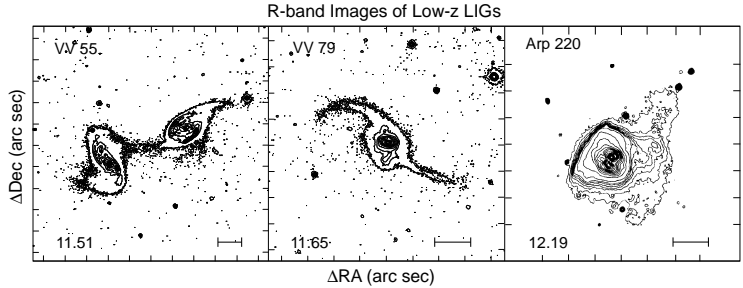, width=10cm}}
\caption{A representative subsample of R-band images of LIGs (Mazzarella et al.
1999) 
from the {\it IRAS\/} RBGS (Sanders et al. 1999), illustrating
the strong interactions/mergers that are characteristic of nearly all objects 
with $L_{\rm ir} > 10^{11.5}{\ts}L_\odot$.  The scale bar represents 10{\ts}kpc,
tick marks are at 20$^{\prime\prime}$ intervals, and 
log{\ts}($L_{\rm ir}/L_\odot$) is indicated in the lower left of each panel.}
\end{figure}

\vspace{-0.5cm}
\subsection{High-z SCUBA sources ($z${\ts}$\sim$1--5)}

Progress in identifying optical/near-IR counterparts of the SCUBA
deep-field sources has been frustratingly slow, due in large part
to the intrinsic faintness of optical/near-IR counterparts.  
However, this is now understood -- from far-UV studies of local ULIGs 
(e.g. Trentham, Kormendy \& Sanders 1999) -- as what would be expected for 
ULIGs at $z${\ts}$>${\ts}1.  Local ULIGs, if placed at $z${\ts}$=$1--4, would
have 
apparent magnitudes in the range $m_{\rm B}${\ts}$\sim${\ts}27--32,  
$m_{\rm I}${\ts}$\sim${\ts}25--29, and $m_{\rm K}${\ts}$\sim${\ts}21--24{\ts}!

Currently only $\sim${\ts}25{\ts}\% 
of the sources with $S_{850}${\ts}$>${\ts}3{\ts}mJy appear to have ``secure'' 
identifications and redshifts (e.g. Barger et al. 1999). 
However, the best studied of these 
have properties (e.g. magnitudes, morphology, spectra, gas-content) 
similar both to local ULIGs as well as to the small sample of high-$z$ 
ULIGs discovered in the {\it IRAS} faint source database 
(see Scoville, these proceedings). 

\vspace{-0.25cm}
\section{Summary: ``Star Formation History" of the Universe}

\begin{figure}[ht]
\vspace{-0.25cm}
\centerline{\epsfig{file=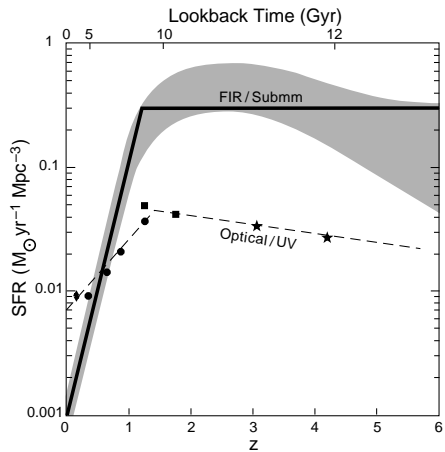, width=6cm}}
\vspace{-.75cm}
\caption{The ``star formation rate'' vs. $z$ for optical/UV 
and far-IR/submm selected galaxies. 
($H_{\rm o} = 50${\ts}km{\ts}s$^{-1}$Mpc$^{-1}$,
$q_{\rm o} = 0.5$ is used for consistency with previously published 
optical versions of this plot.).
In the optical/near-UV, the mean co-moving SFR 
is determined from the total observed rest-frame
UV luminosity density of galaxies (solid diamond: Trayer \etal\  1998;
solid circles: Cowie, Songalia \& Barger 1999; 
solid squares: Connolly \etal\  1997; 
solid stars: Steidel \etal\  1999).  
The shaded
region and thick solid line represent the maximum contribution to the SFR from
far-IR/submm sources (i.e. assuming all of the far-IR/submm
emission is powered by young stars) using models with a range of
$z$-distributions which are consistent with both the current
observations of 850{\ts}$\mu$m SCUBA sources (Blain \etal\  1999; Barger et al.
1999), 
and the local volume density of LIGs  
(Sanders et al. 1999).}
\vspace{-2.25cm}
\end{figure}
\vspace{-0.25cm} 
 
What the SCUBA deep surveys now make abundantly clear is that 
a substantial fraction of the ``activity" in galaxies
at high redshifts ($z >${\ts}1) is obscured by dust, and, therefore has
been missed in deep optical/UV surveys.  This is graphically illustrated 
in Figure 6 using the latest SCUBA redshift distribution estimates of Barger et
al. (1999), 
assuming that all of the far-IR/submm luminosity is 
powered by star 

\vspace{-2.0cm}
\noindent
formation, and then comparing with similar plots derived 
for deep optical/UV surveys. Figure 6 suggests that the SCUBA sources 
dominate the {\it observed} optical/UV SFR by
at least a factor of 10 at $z${\ts}$>${\ts}1. 
 
What is the relationship of the
SCUBA sources to the optically selected high-$z$ population of starburst
galaxies ?
One view is that the SCUBA sources are indeed just the most heavily reddened
objects 
already contained in the optical samples.  Favoring this view is the
evidence (summarized by Steidel \etal\  1999) that on average the more
luminous objects in optical samples are also redder, such that after correction 
for extinction (typically by a mean factor of $\sim${\ts}3--5) using 
models developed for nearby starburst galaxies 
(e.g.  Meurer \etal\  1997;
Calzetti 1997) they would have intrinsic luminosities equivalent to that of
the SCUBA sources (i.e. $\gsim 10^{12} L_\odot$).
However, there is little current evidence to show that the SCUBA detections
are related to the most heavily reddened optical sources, or that applying a
mean 
dust correction to all optical/UV sources is advised. 

An alternative view is that the SCUBA sources represent an inherently distinct
population, for example the formation of spheroids and massive
black holes, both of which are triggered by the merger of two large gas-rich
disks 
(e.g. Kormendy \& Sanders 1992; Kormendy \& Richstone 1995; SM96). 
Favoring this view is the fact that the strong evolution 
for {\it IRAS} ULIGs and SCUBA sources 
at $z${\ts}$<${\ts}1, and a possible peak in the range $z${\ts}$\sim${\ts}1--3, 
is similar to what is observed for QSOs
(e.g. Schmidt \etal\  1995) and radio
galaxies
(Dunlop 1997).  For the UV/starburst population, the more gradual decrease 
at $z${\ts}$<${\ts}1 and the flat redshift distribution 
at $z${\ts}$>${\ts}1 (Steidel \etal\  1999) might better
 represent
the building of disks over a wider range of cosmic time.


\end{article}
\end{document}